\documentclass[aps,nofootinbib,twocolumn,showpacs,superscriptaddress,amssymb]{revtex4-1}
\date{\today}
\usepackage{graphicx}
\usepackage{color}
\begin{document}

\title{Abrupt changes in alpha decay systematics as a manifestation
of collective nuclear modes}

\author{C. Qi}
\email{chongq@kth.se}
\affiliation{KTH, Alba Nova University Center,
SE-10691 Stockholm, Sweden}

\author{A. N. Andreyev}
\affiliation{Instituut voor Kern-en Stralingsfysica, K. U. Leuven, B-3001 Leuven, Belgium}
\affiliation{School of Engineering and Science, University of the West of Scotland, Paisley PA1 2BE, United Kingdom}

\author{M. Huyse}
\affiliation{Instituut voor Kern-en Stralingsfysica, K. U. Leuven, B-3001 Leuven, Belgium}

\author{R. J. Liotta}
\affiliation{KTH, Alba Nova University Center,
SE-10691 Stockholm, Sweden}

\author{P. Van Duppen}
\affiliation{Instituut voor Kern-en Stralingsfysica, K. U. Leuven, B-3001 Leuven, Belgium}

\author{R. A. Wyss}
\affiliation{KTH, Alba Nova University Center, SE-10691 Stockholm,
Sweden}

\begin{abstract}
An abrupt change in $\alpha$ decay systematics around the $N=126$ neutron shell closure is discussed.
It is explained as
a sudden hindrance of the clustering of the nucleons that eventually form
the $\alpha$ particle. This is because the clustering induced by the pairing
mode acting upon the four nucleons is inhibited if
the configuration space does not allow a proper manifestation
of the pairing collectivity.
\end{abstract}
\pacs{21.10.Tg, 23.60.+e, 27.80.+w, 21.60.Cs}
\maketitle
\section{Introduction}
\label{intro}

It is nearly a century ago that the Geiger-Nuttall law, which was to
revolutionize physics by its implications, was formulated based on $\alpha$ decay systematics~\cite{gn,gn2}.
Indeed, its explanation by Gamow~\cite{Gam28} and also by Gurney and Condon~\cite{gc} required to accept the
probabilistic interpretation of Quantum Mechanics. The extend to which this was
revolutionary can perhaps best be gauged by noticing the multitude of
models that have been put forward by outstanding physicists as an alternative
to the probabilistic interpretation. This debate rages even at present~\cite{hoo}.

The Gamow theory reproduced the Geiger-Nuttall law nicely. One can
assert that this is an effective theory, where concepts like ``frequency of
escape attempts" have to be introduced. Yet Gamow's theory is so
successful that even today it is applied, with minor changes, in the studies
of radioactive decays (e.g., Refs.~\cite{arun,zhang,Ren09}). In fact, a proper
calculation of the decay process needs to address first the
clustering of the nucleons at a certain distance outside the nuclear surface and, in a
second step, the evaluation of the penetrability through the Coulomb
and centrifugal barriers
should be performed at the distance where
the cluster was formed. The first step is a challenging
undertaking because a proper description of the cluster in terms of its
components
requires a microscopic many-body framework that is very complicated. This is the reason why
usually effective approaches are used when dealing with clusterization. That is, one
evaluates the penetrability, which is an easy task specially if semiclassical
approaches are applied, and free parameters are introduced for the clustering process trying to
reproduce experimental data.

One may then wonder why effective approaches have been so successful. The reason is
that the $\alpha$-particle formation probability usually varies from
nucleus to nucleus much less than the penetrability. In
the logarithm scale of the Geiger-Nuttall law the differences in the
formation probabilities are usually small fluctuations along the straight
lines predicted by that law~\cite{Buck90} for different
isotopic chains.
The importance of a proper treatment of $\alpha$ decay was attested by a
recent calculation which shows that the different lines can be merged in a
single line. One thus obtained a generalization of the Geiger-Nuttall law
which holds for all isotopic chains and all cluster radioactivities~\cite{qi08,qi09}.
In this universal decay law (UDL) the penetrability is
still a dominant quantity. By using three free parameters only, one finds
that all known ground-state to ground-state radioactive decays are explained
rather well.
This good agreement is a consequence of the smooth transition in the nuclear
structure that is often found when going from a nucleus to its neighboring
nuclei. This is also the reason why, e.g., the BCS approximation works so
well in many nuclear regions.

In this paper we will show that, when a sudden transition occurs in a given
chain of nuclei,
departures from the UDL can be seen. Our aim is to understand why this
difference appears. We will also try to discern whether one can, in general,
obtain information about the structure of the nuclei involved in the decay.
This would be an important task because many regions of the nuclidic chart now under
scrutiny,
especially superheavy nuclei, are radioactive and often $\alpha$ decay is the
only tool that one has to explore their structure.

In Section~\ref{fampl} the formation amplitude is defined. In Section~\ref{expd}
$\alpha$ formation amplitudes extracted from experimental data are presented and
abrupt changes are noted. In Section~\ref{eval} the
evaluation of the formation amplitudes and half-lives of Po isotopes,
which do not follow the UDL, is
performed. A summary and conclusions are in Section~\ref{sumcon}.

\section{The formation amplitude}
\label{fampl}

After the seminal Gamow's paper, the first attempt to formulate a proper
treatment of $\alpha$ decay was based on the compound system
theory developed by Teichmann and Wigner~\cite{wig}. Here the very
complicated process occurring as the compound system decays,
is divided into an ``internal region",
where the compound state is restricted,
and the complementary ``external region". This division is such that in the
external region only the Coulomb and centrifugal forces are important. Thus
the decaying system behaves like a two-particle system. This formulation
was applied by Thomas to $\alpha$-decay~\cite{Tho54} to obtain the classical
expression for the decay width $\Gamma_l$ as
\begin{equation}\label{Tho}
\Gamma_l(R)=2{\cal P}_l(R) \frac{\hbar^2}{2\mu R} |{\cal F}_l(R)|^2,
\end{equation}
where $l$ is the angular momentum carried by the outgoing $\alpha$-particle,
${\cal P}$ is the penetration probability and
$\mu$ is the reduced mass corresponding to the final system consisting of an $\alpha$ particle and a daughter nucleus.
$R$ is the radius dividing the internal and external regions. At this
point the wave function of the $\alpha$-particle already formed in the
internal region is matched with the corresponding outgoing
two-body wave function in the external region. The amplitude of the wave
function in the internal region is the formation amplitude, i.e.,
\begin{equation}\label{foram}
{\cal F}_l(R)=\int d{\mathbf R} d\xi_d d\xi_\alpha
[\Psi(\xi_d)\phi(\xi_\alpha)Y_l(\mathbf R)]^*_{J_mM_m}
\Psi_m(\xi_d,\xi_\alpha,\mathbf{R}),
\end{equation}
where $d$, $\alpha$ and $m$ label the daughter, $\alpha$ particle and mother
nuclei, respectively. $\Psi$ are the intrinsic wave functions and $\xi$ the corresponding
intrinsic coordinates. $\phi(\xi_\alpha)$ is a Gaussian function of the relative
coordinates of the two neutrons and two protons that constitute the $\alpha$-particle, coupled
to zero angular momentum~\cite{Mang60,Ton79}. The rest of the notation is standard.

One sees from Eq.~(\ref{foram}) that $\mathcal{F}_l(R)$ would indeed be the
wave function of the outgoing $\alpha$ particle $\psi_\alpha(R)$ if the
mother nucleus would behave at the point $R$ as
\begin{equation}\label{mother}
\Psi_m(\xi_d,\xi_\alpha,\mathbf{R})=
[\Psi(\xi_d)\phi(\xi_\alpha)\psi_\alpha(R) Y_l(\mathbf R)]_{J_mM_m}.
\end{equation}
Since this is usually a smal component of the mother nucleus wave function,
the corresponding formation amplitude (\ref{foram}) is small, of the order of 
$10^{-2}$ \cite{Lovas98}. The main problem in the evaluation of this
quantity is the description of the clusterization
of the four nucleons that eventually become the $\alpha$-particle. In
pursuing this task one has found that the mode that determines clusterization 
is the pairing vibration~\cite{Ton79,Jan83}. In fact, the study of 
$\alpha$-clusterization gave rise
to the realization that there should be a giant pairing vibration lying high
in the nuclear spectra~\cite{Her85,Dus09}.
It is also interesting to notice that the $\alpha$-clusterization in $\alpha$-decaying
nuclei has triggered the appearance of effective models where the wave function
of nuclei such as $^{212}$Po is assumed to have the form (\ref{mother}). The
spectra thus obtained agree well with the corresponding experimental data~\cite{Buc95}.

Going back to Eq.~(\ref{Tho}), the wave function corresponding
to the external region, i.e., to the outgoing
channel, gives rise to the penetration probability
${\cal P}_l(R)=kR/(G_l^2+F_l^2)$, where $G_l$ and $F_l$ are the irregular and
regular Coulomb functions, respectively. From Eq.~(\ref{Tho}) it is
straightforward to see that the width $\Gamma_l(R)$ cannot depend upon R,
since outside the range of the nuclear interaction (i.e., just outside the
nuclear surface) the internal and external wave functions
are the same~\cite{enr}, i.e.,  $\mathcal{F}_l(R)\propto G_L(R)+iF_l(R)$.
This is of course valid provided that the formation amplitude was evaluated
properly. In fact a way of probing the calculation is just by investigating
whether the width is dependent upon $R$, and in such a 
case by how much~\cite{Dod89}.

The $\alpha$-decay half-life can be written as
\begin{equation}\label{life}
T_{1/2}=\frac{\hbar\ln2}{\Gamma_l} = \frac{\ln2}{\nu} \left|
\frac{H_l^+(\chi,\rho)}{R \mathcal{F}_l(R)} \right|^2,
\end{equation}
where $\nu$ is the outgoing velocity of the emitted particle.
The distance $R$  will be taken as the touching point, i.e.,
$R=R_0(A_d^{1/3}+4^{1/3})$, with $R_0$=1.2 fm.
The other quantities are standard, i.e., $H_l^+(\chi,\rho)$ is the
Coulomb-Hankel function with arguments $\chi=4Z_d e^2/\hbar \nu$ and
$\rho=\mu\nu R/\hbar$. 

In microscopic theories the formation amplitude is evaluated
starting from the single-particle degrees of freedom of the neutrons
and protons that eventually become the cluster.
This requires advanced computing facilities as well
as suitable theoretical schemes to describe the clustering
process. It is therefore not surprising that the first calculations
of absolute decay widths were performed after the appearance of the
shell model. These calculations had limited success due to the small
shell model spaces that could be included at that time~\cite{Mang60}. Yet, in
retrospect it is surprising to note the deep insight the pioneers in these
shell model calculations had on the role of
configuration mixing to induce clustering~\cite{ras65}. That this was indeed
the case was shown much later~\cite{Jan83,Ton79} in the case of the decay
of the nucleus $^{212}$Po with two protons and two neutrons outside the
doubly magic core, $^{208}$Pb, which has been considered as a textbook example
in illustrating the clustering and decay of the alpha particle in heavy nuclei (see, e.g., Ref.~\cite{Ast10} and references therein). In fact this case is very important for the present paper,
since the most significant departure of the UDL from experimental data that we will investigate is in
the ground-state to ground-state decays of Po isotopes.

\section{The experimental data}
\label{expd}

\begin{figure}[htdp]
\includegraphics[scale=0.45]{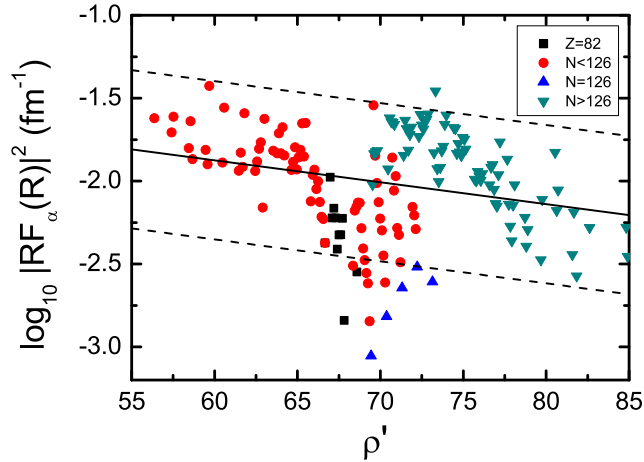}\\
\caption{(Color online) $\log_{10}|RF(R)|^2$ as a function
of $\rho'$. The solid line denotes the smooth behavior of the
UDL. The values between the two dashed lines differ from the corresponding
UDL values by a factor of at most three.}\label{fvsrp}
\end{figure}

Using the experimental decay half-lives~\cite{Audi03} 
one can extract the formations amplitudes by applying 
Eq.~(\ref{life}). One thus obtains
\begin{equation}
\label{expfor}
\log_{10} |R{\cal F}_{\alpha}(R)|=-\frac{1}{2}\log_{10} T^{{\rm
Expt.}}_{1/2} +\frac{1}{2} \log_{10} \left[ \frac{\ln
2}{\nu}|H^+_0(\chi,\rho)|^2\right].
\end{equation}
This is shown in Fig.~\ref{fvsrp} for different even-even isotopes
 as a function of
the quantity  $\rho' = \sqrt{2\mathcal{A}Z_d(A_d^{1/3}+4^{1/3})}$, where
$\mathcal{A}=4A_d/A_m$. This is one of the two variables that defines the UDL~\cite{qi08}.


One notices in Fig.~\ref{fvsrp} that at
$\rho'\approx 70$ a division occurs between decays corresponding to $N<
126$ and $N>126$. Perhaps even more important is that
for most cases the UDL predicts the experimental values within a factor of
three, except for $N=126$, where the difference becomes about one order of 
magnitude. This is so distinct that one may even
suspect that the difference in the values of ${\cal F}_{\alpha}$ when going 
from one nucleus to its neighbors in the vicinity of $N=126$ overruns the
corresponding differences in the penetrability.
If one understands the reason of this large variation, $\alpha$ decay may
provide a powerful tool to study the
structure of decaying nuclei. This point will be analyzed in the next Section.

The case that shows the most significant hindrance corresponds to the $\alpha$ 
decay of the nucleus $^{210}$Po, with $\log_{10} |R{\cal F}_{\alpha}(R)|^2<-3~{\rm fm}^{-1}$. The symbols with  $\log_{10} |R{\cal F}_{\alpha}(R)|^2\sim-2.7~{\rm fm}^{-1}$ correspond to the $\alpha$ decays of nuclei $^{208}$Po ($N=124$), $^{212}$Rn ($N=126$) and $^{194}$Pb.

Finally, it is worthwhile to point out that this sudden change in
$\alpha$-decay systematics at $N=126$ has also been noticed in 
Refs.~\cite{Bata88,Toth99,Wau93,Van03}. Moreover, in the semiclassical approaches of Refs.~~\cite{Brown92,Poe07,Poe83,Poe91,Poe06a,Poe06b,Ism10}
the decay half-lives of nuclei with $N=126$ are significantly underestimated. In the alpha-decay formula of Refs.~\cite{Poe06b,Poe07}, an empirical correction term has been introduced to taken into account the large underestimation around shell closures.

\section{Evaluation of the half-lives in Po isotopes.}
\label{eval}

In this Section we will analyze, within a microscopic formalism, the half-lives of the isotopes that show the kink at $N=126$ discussed above. We will take the decay of $^{210}$Po as a typical example and compare it with that of $^{212}$Po.
To compare with experimental data, we extract the magnitude of the
formation amplitudes from measured half-lives by using
Eq.~(\ref{expfor}). One thus obtains the value
${\cal F}_{\alpha}(R)=3.305\times10^{-3}$~fm$^{-3/2}$ in $^{210}$Po and
${\cal F}_{\alpha}(R)=1.082\times10^{-2}$~fm$^{-3/2}$
in $^{212}$Po, where we used $R=9.0$~fm.
These correspond to a variation in the formation
amplitudes by a factor of 3.28, that is a factor of 10.73
in the formation probabilities.

Within the shell model a four-particle state $\alpha_4$ in $^{212}$Po can be 
written as 
\begin{equation}
\label{msmwf}
|^{212}{\rm Po}(\alpha_4)\rangle=\sum_{\alpha_2 \beta_2} X(\alpha_2\beta_2;\alpha_4) 
|^{210}{\rm Pb}(\alpha_2)\otimes^{210}{\rm Po}(\beta_2)\rangle
\end{equation} 
where $\alpha_2$ ($\beta_2$) labels two-neutron (two-proton) states. 
For the ground state of $^{212}$Po, it was found~\cite{Dod89} that $X(^{210}{\rm Pb(gs)}\otimes~^{210}{\rm Po}{\rm (gs)})=0.9$, while
$X(^{210}{\rm Pb}(2^+_1)\otimes~^{210}{\rm Po}(2^+_1))=-0.3$. 

Each of the terms in Eq.~(\ref{msmwf}) corresponds to neutron-neutron (nn) or proton-proton (pp) states, 
i.e., states determined by the nn or pp interaction. The neutron-proton (np)
interaction mixes those states. In other words, the amplitudes $X$ are
influenced by the np interaction. If this interaction is neglected, then
only one of the configurations in Eq.~(\ref{msmwf}) would appear. This
is done, for instance, in cases where the correlated four-particle
state is assumed to be provided by collective vibrational states.
Rather typical examples of such states are $|^{210}{\rm Pb(gs)}\rangle$ and 
$|^{210}{\rm Po(gs)}\rangle$. It is therefore not surprising that
calculations have been performed by assuming that $|^{212}{\rm Po(gs)}\rangle$ 
is a double pairing vibration~\cite{Ton79,Jan82}, i.e.,
\begin{equation}
\label{paiwf}
|^{212}{\rm Po(gs)}\rangle=|^{210}{\rm Pb(gs)}\otimes~^{210}{\rm Po(gs)}\rangle.
\end{equation}
The corresponding formation amplitude acquires the form,
\begin{eqnarray}\label{po212}
&&{\cal F}_{\alpha}(R;^{212}{\rm Po(gs)}) =
\int d\mathbf{R}
d\xi_{\alpha}\phi_{\alpha}(\xi_{\alpha})
\nonumber
\\&&\times \Psi(\mathbf{r_1},\mathbf{r_2};^{210}{\rm Pb(gs)})
\Psi(\mathbf{r_3},\mathbf{r_4};^{210}{\rm Po(gs)}),
\end{eqnarray}
where $\mathbf{r_1},\mathbf{r_2}$ ($\mathbf{r_3},\mathbf{r_4}$)
are the neutron (proton) coordinates and $\mathbf{R}$ is the center
of mass of the $\alpha$ particle.

With this expression for the formation amplitude the experimental 
half-life is reproduced rather well if a large number of
high-lying configurations is included. These configurations are needed
to describe the clusterization between the two neutrons and the two protons 
in the $\alpha$ particle. Yet the 
corresponding $\alpha$-decay half-life is still too small
by more than one order of magnitude. This is because the neutron-proton 
interaction is not included in Eq.~(\ref{paiwf}). When this is done,
and again a large configurations space is used, the neutrons and protons
become also clustered, 
enhancing the value of the half-life. It is also important to underline
that the inclusion of the large configuration space provides a half-life
which is independent upon the matching point $R$~\cite{Dod89}.

We reproduced these calculations by using a surface delta interaction and
nine major shells of a harmonic oscillator (HO) representation. 
The decay of the
nucleus $^{210}$Po(gs) leads to the daughter nucleus $^{206}$Pb(gs), which
is a two-hole state. Here we used the five HO major shells corresponding
to the single-hole states that describe the wave function of 
$^{206}$Pb(gs) as
\begin{eqnarray}
\label{twoh}
&&|^{206}{\rm Pb(gs)}\rangle = \sum_{h_1\leq h_2} X(h_1h_2;^{206}{\rm Pb(gs)})
\nonumber
\\&&\times
\frac{(b^+_{h_1}b^+_{h_2})_{0^+}}{\sqrt{2}} |^{208}{\rm Pb(gs)}\rangle,
\end{eqnarray}
where  $h$ labels single-hole states and the hole creation operator is standard,
i.e., $b^+_{jm}=(-1)^{j-m}c_{j-m}$. The formation amplitude becomes,
\begin{eqnarray}\label{po210}
&&{\cal F}_{\alpha}(R;^{210}{\rm Po(gs)}) = \int d\mathbf{R}
d\xi_{\alpha}\phi_{\alpha}(\xi_{\alpha})
\nonumber
\\&&\times
\Psi^*(\mathbf{r_1},\mathbf{r_2};^{206}{\rm Pb(gs)})
(\mathbf{r_3},\mathbf{r_4};^{210}{\rm Po(gs)}).
\end{eqnarray}

By comparing Eqs.~(\ref{po212}) and (\ref{po210}) one sees that the only
difference between the two expressions is the two-neutron wave function, which
corresponds to the two-particle state $^{210}$Pb(gs) in Eq. (\ref{po212})
and to the two-hole state $^{206}$Pb(gs) in Eq. (\ref{po210}). Therefore the
kink observed experimentally should be related to the difference in
clusterization induced by the pairing force in these two cases.
To analyze the clustering features we will consider only the spin-singlet component, i.e.,
$(\chi_1\chi_2)_0$, of
the two-body wave function, since that is the only part entering the
intrinsic $\alpha$-particle wave function. This component has the form,
\begin{eqnarray}
\label{sinwf}
&&\Psi_2(r_1,r_2;\theta_{12})=\frac{1}{4\pi}
\sum_{p\leq q}\sqrt{\frac{2j_p+1}{2}} X(pq;{\rm gs})
\nonumber \\&&
\times 
\varphi_p(r_1)\varphi_q(r_2)P_{l_p}(\cos\theta_{12}),
\end{eqnarray}
where $\varphi$ is the single-particle wave function and $P_l$ is the Legendre polynomial
of order $l$ satisfying $P_l(\cos0)=1$ (notice that for
the ground states 
studied here it is $l_p=l_q$). As mentioned above, 
the pairing vibrations show strong clustering features as the number of
single-particle states is increased~\cite{Jan83}.
But another manifestation of the pairing collectivity is
an enhancement of the wave function on the nuclear surface. The reason of this
enhancement is that all configurations
contribute with the same phase in the building up of the two-particle wave
function on the nuclear surface. 
The same mechanism increases the $\alpha$ formation amplitude and,
therefore, the relative values of the wave functions of
$^{210}$Pb(gs), $^{210}$Po(gs) and $^{206}$Pb(gs) on
the nuclear surface
give a measure of the importance of the
corresponding formation amplitudes. 

\begin{figure}[htdp]
\includegraphics[scale=0.45]{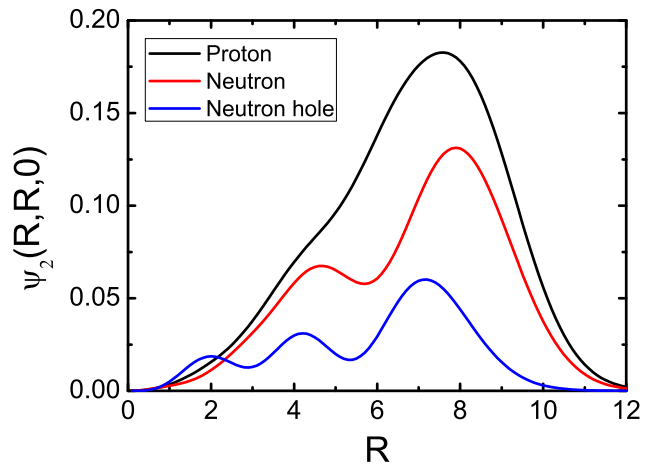}\\
\caption{(Color online) The two-body wave function $\Psi_{2}$(R,R,0)
corresponding to the pairing vibrations in the two-neutron particle
$^{210}$Pb(gs), two-proton particle $^{210}$Po(gs) and two-neutron
hole $^{206}$Pb(gs) cases.}
\label{wfvsr}
\end{figure}

To study the behavior of the two-particle wave functions  
we will apply Eq. (\ref{sinwf}) with 
$r_1=r_2$ and $\theta_{12}=0$. This is reasonable since
due to clustering the
wave function is strongly peaked at $\theta_{12}=0$.
Calling $R=r_1$, we have plotted   
in Fig.~\ref{wfvsr} $\Psi_2(R,R,0)$ as a function of $R$.
One sees that the wave functions are
indeed strongly enhanced at the nuclear surface, as expected. But the
important feature for us is that the enhancement is strongest in
$^{210}$Pb(gs) and weakest in $^{206}$Pb(gs).
This is because there is a relatively small number of configurations
in the hole-hole case. In addition, the radial wave functions corresponding
to the high-lying particle states
extend farther out in space with respect to the hole configurations.  

\begin{figure}[htdp]
\includegraphics[scale=0.45]{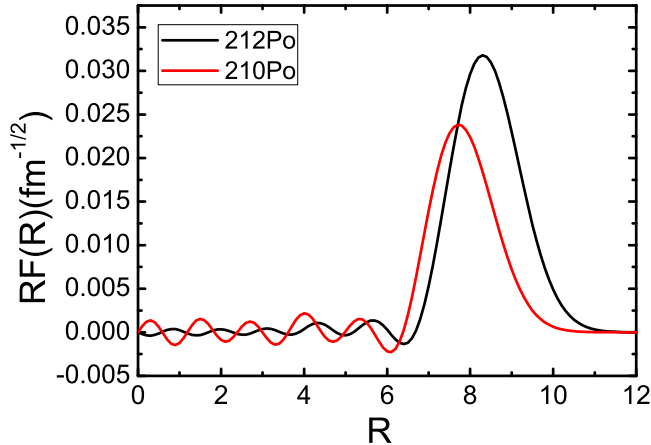}\\
\caption{(Color online) The $\alpha$ formation amplitudes
$R\mathcal{F}_{\alpha}(R)$ corresponding to the nuclei 
$^{212}$Po(gs) and $^{210}$Po(gs).}
\label{form}
\end{figure}

With these two-body wave functions we proceeded to evaluate the $\alpha$ formation amplitudes
in $^{212}$Po(gs) and $^{210}$Po(gs). The results are shown in Fig.~\ref{form}.
One finds from this figure that with $R=9$~fm the observed ratio
between the formation amplitudes in $^{212}$Po and $^{210}$Po can be
reproduced nicely.

\subsection{The neutron-proton interaction}
\label{pni}

We have assumed (Eq.~(\ref{paiwf})) that $^{212}$Po(gs) is virtually a correlated two-neutron
two-proton state. The same is valid for  $^{210}$Po(gs), although here the state
is a correlated two-particle (proton) two-hole (neutron) state. This is a manifestation of 
the pairing vibrational character of two-particle 
states in the Pb region. That is, the correlated two-particle and  
two-hole states in the Pb region can be considered as boson degrees of freedom. This was one of
the main assumptions in the Nuclear Field Theory~\cite{NFT} as well as in the original Interacting Boson 
Model~\cite{IBM}. This assumption implies that the neutron-proton 
interaction does not play a very important role in the spectroscopy of the
states. However, as we have seen, this interaction induces the
clusterization of neutrons and protons. As pointed out in Ref.~\cite{Ton79},
in the Pb region 
low lying neutron and proton single-particle states are very different 
from each other, or are particle-hole states. Therefore the neutron-proton
interaction affects only slightly the ground states and the clusterization 
occurs through high-lying configurations. This point is supported by our 
shell-model calculations with the surface delta as well as realistic interactions.

Only when neutrons and protons move in the same orbits it is expected that
the neutron-proton interaction would affect significantly the spectroscopic properties as well
as the clusterization. We confirmed this by studying 
a model case in which the core consists of an equal number of neutrons and
protons, namely the $\alpha$ decay of the fictional nucleus
$^{168}$Po(gs), with two neutrons and two protons outside the core
$^{164}$Pb(gs). We used, for neutrons as well as protons, the single-particle 
states corresponding to protons in the study performed above for $^{210}$Po.
We also used the same
interaction. As expected, we again found that neutrons and 
protons are strongly clustered as a result of the corresponding 
pairing interaction, But also the 
proton-neutron clustering is  
significantly enhanced by the proton-neutron interaction,
This indicates that in realistic $N=Z$ nuclear regions, for instance around
$^{100}$Sn, there should be a large probability to form an alpha particle (see, e.g., Ref.~\cite{Lid06}).
One can thus conclude that  alpha decay probes may be a powerful tool to 
get information about the structure of heavy $N\approx Z$ nuclei which, 
otherwise, would be difficult to reach.

\section{Summary and Conclusions}
\label{sumcon}

In this paper we have applied the recently proposed  universal decay 
law (UDL)~\cite{qi08}
to perform a systematic calculation of $\alpha$ decay 
half-lives over all experimentally known cases.
We found that although the UDL reproduces nicely most available
experimental data, as expected, there is a case where it fails by a
large factor. This corresponds to the $\alpha$ decays of nuclei with 
neutron numbers equal to or just below $N=126$. 
The reason for this large discrepancy is that in $N\leq126$ nuclei 
the $\alpha$ formation amplitudes
are much smaller than the average quantity predicted by the UDL 
(Fig.~\ref{fvsrp}). This is an indication 
that the $\alpha$ decay transitions in these nuclei are hindered with respect 
to those in the open-shell region. 

The case that shows the most significant hindrance corresponds to the $\alpha$ 
decay of the nucleus $^{210}$Po
for which standard shell-model calculation is feasible. Starting from the formal 
definition of Eq.~(\ref{foram}),
we calculated the $\alpha$ formation amplitude of $^{210}$Po and compared it with that of $^{212}$Po. 
In these two cases the formation amplitudes can be described by
the simple expressions (\ref{po212}) and (\ref{po210}).
We found that the formation amplitude in $^{210}$Po is hindered with respect
to the one in $^{212}$Po due to the hole character of the neutron states in
the first case. This is a manifestation of the mechanism that induces
clusterization, which is favored by the presence of high-lying
configurations. Such configurations are more accessible in the
neutron-particle case of $^{212}$Po than in the neutron-hole case of
$^{210}$Po. This is a general feature in nuclei where neutrons and protons
occupy different low-lying major shells. 
If instead both types of particles occupy the same shells, the
neutron-proton  
interaction is very effective to induce clustering and the formation
amplitude increases strongly. This was the case in a calculation that we
performed considering the fictitious $N=Z=84$ $^{168}$Po isotope as    
the mother nucleus, indicating that even in physically meaningful $N=Z$ 
nuclear region $\alpha$ decay can be enhanced by large factors.

This allows one to assert that 
$\alpha$ decay is a powerful tool to investigate the shell structure of 
very unstable nuclei (including superheavy ones),
where often only $\alpha$-decay quantities can be measured.

\section*{Acknowledgments}
This work has been supported by the Swedish Research Council (VR), FWO-Vlaanderen (Belgium), GOA/2004/03 (BOF-K.U.Leuven), the IUAP-Belgian State-Belgian Science Policy-(BriX network P6/23), and by the European Commission within the Sixth Framework Programme
through I3-EURONS (Contract RII3-CT-2004-506065).

\end{document}